\documentclass[aps,pra,groupedaddress,notitlepage,nofootinbib,superscriptaddress,longbibliography]{revtex4-1}
\usepackage{amsmath}
\usepackage{amssymb,bm}
\usepackage{amsthm,color,xcolor,dsfont}
\usepackage{graphicx} 
\usepackage{caption}
\usepackage[colorlinks=true, linkcolor=blue, urlcolor=blue, citecolor=blue, hyperindex, breaklinks]{hyperref}
\usepackage{ulem}
\newcommand{\ket}[1]{|#1\rangle} 
\newcommand{\bra}[1]{\left\langle #1 \right|} 
\newcommand{\kb}[2]{\ket{#1}\bra{#2}} 
 
\newtheorem*{definition*}{Definition}

\begin{document}

\title{Comparison of memory thresholds for planar qudit geometries}

\author{Jacob Marks}
\email[Electronic address: ]{jacob.marks@yale.edu}
\affiliation{Department of Physics, Yale University, New Haven, CT, 06520, USA}
\affiliation{Institute for Quantum Computing, University of Waterloo, 
Waterloo, ON, N2L 3G1, Canada}

\author{Tomas Jochym-O'Connor}
\email[Electronic address: ]{tjoc@caltech.edu}
\affiliation{Institute for Quantum Computing, University of Waterloo, 
Waterloo, ON, N2L 3G1, Canada}
\affiliation{Department of Physics and Astronomy, University of Waterloo, 
Waterloo, ON, N2L 3G1, Canada}
\affiliation{Walter Burke Institute for Theoretical Physics, Institute for Quantum Information and Matter, California Institute of Technology, Pasadena, CA, 91125, USA}

\author{Vlad Gheorghiu}
\email[Electronic address: ]{vlad.gheorghiu@uwaterloo.ca}
\affiliation{Institute for Quantum Computing, University of Waterloo, 
Waterloo, ON, N2L 3G1, Canada}
\affiliation{Department of Combinatorics and Optimization, University of Waterloo, 
Waterloo, ON, N2L 3G1, Canada}


\begin{abstract}
We introduce and analyze a new type of decoding algorithm called General Color Clustering~(GCC), based on renormalization group methods, to be used in qudit color codes. The performance of this decoder is analyzed under a generalized bit flip error model, and is used to obtain the first memory threshold estimates for qudit 6-6-6 color codes. The proposed decoder is compared with similar decoding schemes for qudit surface codes as well as the current leading qubit decoders for both sets of codes. We find that, as with surface codes, clustering performs sub-optimally for qubit color codes, giving a threshold of $5.6\%$ compared to the $8.0\%$ obtained through surface projection decoding methods. However, the threshold rate increases by up to $112\%$ for large qudit dimensions, plateauing around~$11.9\%$. All the analysis is performed using \href{https://github.com/jacobmarks/QTop}{QTop}, a new open-source software for simulating and visualizing topological quantum error correcting codes.
\end{abstract}

\maketitle

\section{Introduction\label{sct:intro}}
 \setlength{\parskip}{0cm plus0mm minus0mm}
Quantum error correction (QEC) is of paramount importance for quantum information processing schemes, as any implementation will have imperfections that can lead to loss of coherence. Quantum error correcting codes~(QECC), first introduced by Shor \cite{PhysRevA.52.R2493} two decades ago, seek to address such imperfections in the hopes of ensuring global protection. These codes generalize and extend the notion of classical error correction to both bit and phase flip Pauli errors to protect ``fragile'' quantum states against undesired noise~\cite{PhysRevA.52.R2493, steane1996multiple, PhysRevA.54.1098, quantph.9705052}. Since their inception, a multitude of techniques for constructing good QECC have been developed, such as CSS codes \cite{steane1996multiple, PhysRevA.54.1098}, stabilizer codes \cite{quantph.9705052},  cluster-state based codes \cite{PhysRevLett.86.910}.

	For a physical architecture to implement arbitrary quantum algorithms, it must be able to suppress potential errors that could affect the physical system. Fault-tolerant quantum computation allows for the scalable correction of errors in a controllable manner, and is characterized by the presence of a Threshold Theorem~\cite{Aharonov:1997:FQC:258533.258579, Gottesman.2000, PhysRevLett.98.190504}. Namely, given a QECC, provided that the physical error rate of the gates is below a certain \emph{threshold}, the logical error rate can be made arbitrarily small by extending the code distance. However, computing the threshold for a given code is computationally challenging. Moreover, the value of the threshold will depend on the type of decoder used in the code: once a code develops errors, a classical decoding algorithm must be used to return the code to the codespace. The best decoders are fast and result in the fewest logical errors.
    
Topological codes, introduced by Kitaev \cite{Kitaev20032}, are a subclass of stabilizer codes which make use of topological features in order to protect against local physical errors. These are also among the leading candidates for experimental fault-tolerant implementations. Perhaps the most well-known instance of a topological code is the \emph{surface code}, in which data qubits (two-level quantum systems) are placed on a square lattice and the error correction is performed via measuring appropriate stabilizer generators on a shifted ``dual'' lattice. Another type of topological code, the \emph{color code}~\cite{PhysRevLett.97.180501}, is produced by tiling a surface with three-colorable faces, and associating a distinct variety of stabilizer with each color (usually red, green, and blue). These color codes combine the topological error-protection of the surface code with transversal implementations of Clifford gates, allowing for increased ease in logical computation. While most research thus far has focused on qubits, both surface and color codes can be generalized to $D$-level quantum systems, or \emph{qudits}, which can take on linear superpositions of $D$~distinct values. Indeed, early numerical results suggest that qudits may give better performance by providing more information about the specific set of errors that has occurred~\cite{PhysRevA.87.062338, watson2015fast, anwar2014fast}. 

The first contribution of our work is QTop~\cite{qtop}, a universal numerical framework for simulation and visualization of topological codes of arbitrary code distance, and qudit dimension. Our software includes surface codes, and 6-6-6 color codes - one of the most experimentally-promising semi-regular tilings of the plane, and allows for simulation under arbitrary noise models. Our framework is modular, object-oriented, and simple to use. It can be used to test new decoders, and extending it to 3D~color codes and more exotic topological systems is straightforward. 

The second main contribution is a decoder for qudit color codes. This decoder is inspired by renormalization group clustering techniques prevalent in the qudit surface code decoding literature. We implement the proposed decoder in QTop, and analyze its performance under the generalized bit-flip memory noise model, as the code is a CSS code that can address bit and phase flip errors independently. Memory noise, or errors introduced on the physical data qudits alone (while measurement circuits are perfect), is typically used as a first estimate for the viability of an error correcting code for fault-tolerant computation. We obtain a threshold value of $5.6\%$ in the case of the qubit color code, a drop from $8.0\%$ we obtain using the surface projection method. This type of drop is expected due to the inherently approximate nature of renormalization decoders, and is seen in the case of surface code renormalization decoders~\cite{watson2015fast}. Moreover, as in the case of the surface code, the threshold rate increases with qudit dimension, saturating at a value that is above that of the idealized qubit case, that is~$11.9\%$.

The remainder of this paper is organized as follows. In Sec.~\ref{sct:top} we motivate our study of topological codes, and provide a detailed overview of the two most prevalent types - surface and color codes. In Sec.~\ref{sct:decoding} we formulate the problem of decoding. Sec.~\ref{sct:gsp} describes our novel method for decoding qudit color codes, and in Sec.~\ref{sct:results} we present threshold results for such codes under various error models. Finally in Sec.~\ref{sct:conclusion} we summarize our results, raise some open questions and discuss possible future research directions. 

\section{Topological Codes\label{sct:top}}
\subsection{Motivation}
Topological quantum error correcting codes are regarded as highly promising schemes for fault-tolerant quantum computation as logical states are encoded in highly non-local degrees of freedom of the system. Therefore, in order for physical errors to lead to logical faults, error chains will have to form that will be as non-local as the logical states, unlikely in the event that the noise is not strongly correlated. Moreover, topological codes are characterized by stabilizers that are typically of low weight\footnote{The weight of a tensor product of Pauli operators is defined as the number of non-identity Pauli operators in the product.}, and more importantly, local. Therefore, local errors will lead to local excitations in the stabilizer space, typically allowing for efficient decoding algorithms. In this section, we describe two of the most studied classes of topological codes for the purposes of quantum computing, the surface code and the color code.

\subsection{Surface Codes: Benefits \& Limitations}
The surface code was among the first class of topological codes, proposed by Kitaev~\cite{Kitaev20032}. The surface code is a special instance of the Toric code structure, where smooth and rough boundaries are introduced allowing for the storage of a single logical qubit. Namely, complementary sets of anti-commuting logical operators are represented by excitations that connect differing sets of boundary types, satisfying anti-commutation by intersecting at an odd number of sites in the lattice. 

One of the primary advantages of the surface code is that the stabilizers are given by weight-4 operators, allowing for high threshold values in the case of circuit level noise. However, the surface code is limited in its set of logical transversal gates, that is the set of gates that can be implemented in a bit-wise fashion throughout the code. As such, the surface code faces potential increases in overhead for the implementation of logical gates.

\subsection{Color Codes: Topology \& Transversality}
Color codes were first introduced as an alternative two-dimensional geometrical architecture to the surface code~\cite{PhysRevLett.97.180501}. The color code construction is characterized by having three different boundary types, unlike the surface code, and logical operators are formed by connecting all three boundaries. In this sense, the color code has a symmetry between the boundaries that is not present in the surface code. This symmetry manifests itself in the transversal operators of the code, which we detail below. 

Color codes are CSS codes whose stabilizers, in planar geometries, are given by plaquette operators only. Since all plaquettes hold both $X$~and $Z$~type operators, all plaquettes must intersect at an even number of qubits in order to satisfy commutativity of the stabilizers. The graph is therefore 3-valent and 3-colorable, thus giving rise to the notion of color in the code setting~\cite{PhysRevLett.97.180501, PhysRevA.91.032330}. This condition on the structure of the graph leads to a restricted set of possible configurations for the stabilizers of the code, resulting in possible stabilizer weight distributions of 6-6-6, 4-8-8, and~4-6-12~\cite{landahl2011fault}. Unlike the surface code, there is a symmetry in the two types of stabilizers in the code, resulting in a transversal Hadamard gate. Moreover, the generalized phase gate can be implemented transversally by choosing a careful distribution of positive and negative rotations. As such, all Clifford gates can be implemented transversally for the color code\footnote{The CNOT gate is also transversal since the code is a CSS~code.}, a significant advantage over the surface code. These properties of the color code have allowed recent research in fault-tolerance to develop schemes for alternative methods to magic-state distillation for universal quantum computation~\cite{paetznick2013universal,jochym2014using, bombin2015gauge}. In addition, a recent result by Bombin has shown that three-dimensional color codes can be used in single-shot quantum error correction to circumvent the need to store historical syndrome data when correcting for errors in measurement~\cite{bombin2015single}.

\section{The Problem of Decoding\label{sct:decoding}}

\subsection{Minimum Weight Perfect Matching}

\begin{figure}[h!]
\includegraphics[width=0.27\textwidth]{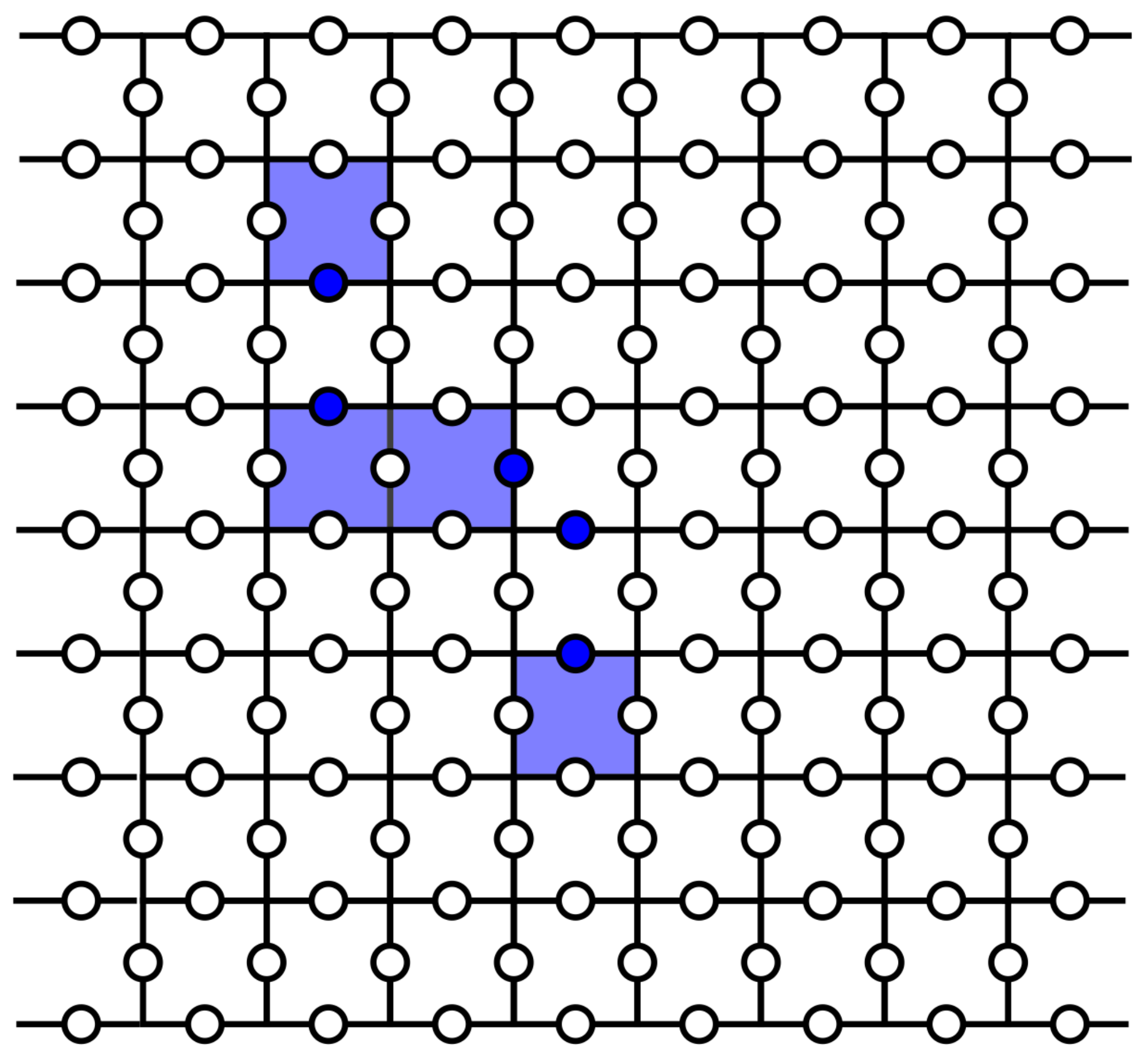}
\caption{Example of a distance~9 implementation of the surface code, where $Z$~stabilizers are defined by weight-4 plaquette operators, while $X$~stabilizers are given by weight-4 star operators. In the presented example, an $X$ error has occurred on the blue colored qubits, causing excitations at the colored plaquettes. These colored plaquettes will serve as the inputs to the decoding algorithm.}
\label{fig:SCPrimal}
\end{figure}

\begin{figure}[h!]
\includegraphics[width=0.25\textwidth]{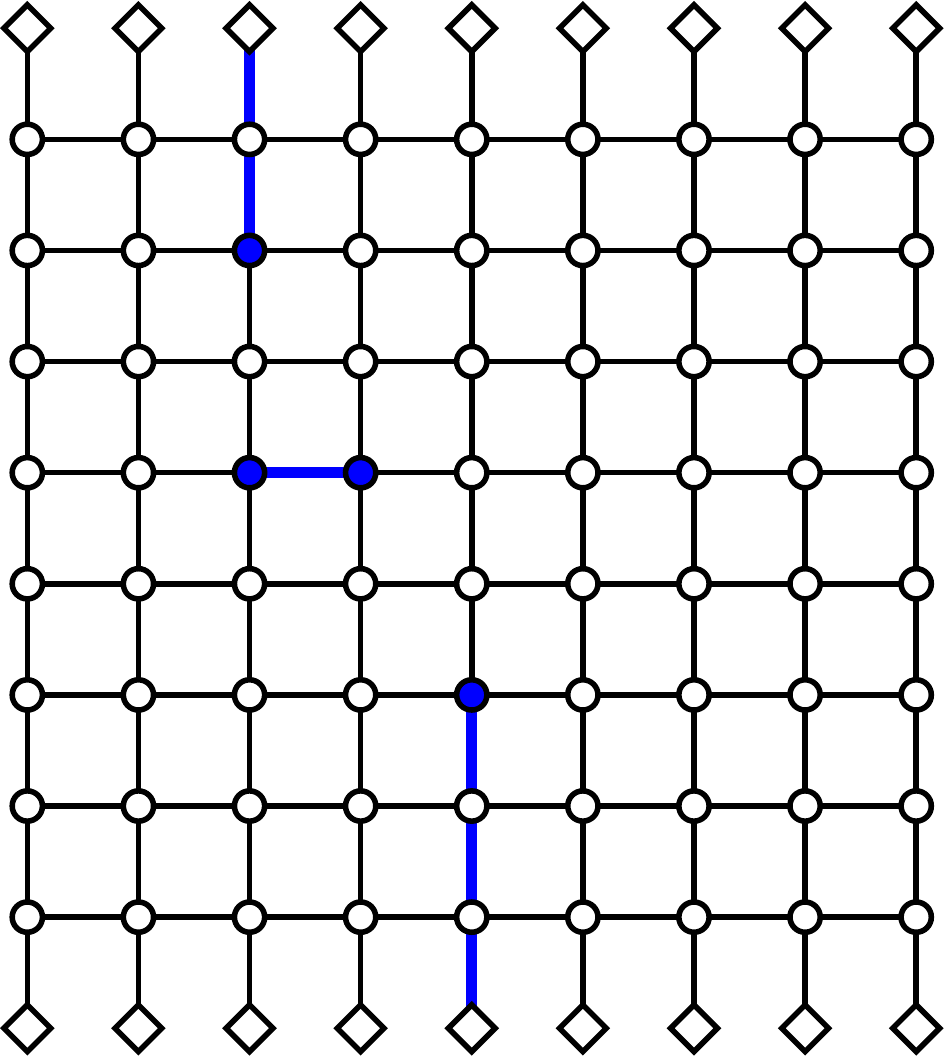}
\caption{Greedy implementation of decoding for the error presented in Fig.~\ref{fig:SCPrimal}. Note that overall the weight of the corrected error chain is higher, even though the two nearest excitations have been paired together.}
\label{fig:SCDualGreedy}
\end{figure}

\begin{figure}[h!]
\includegraphics[width=0.25\textwidth]{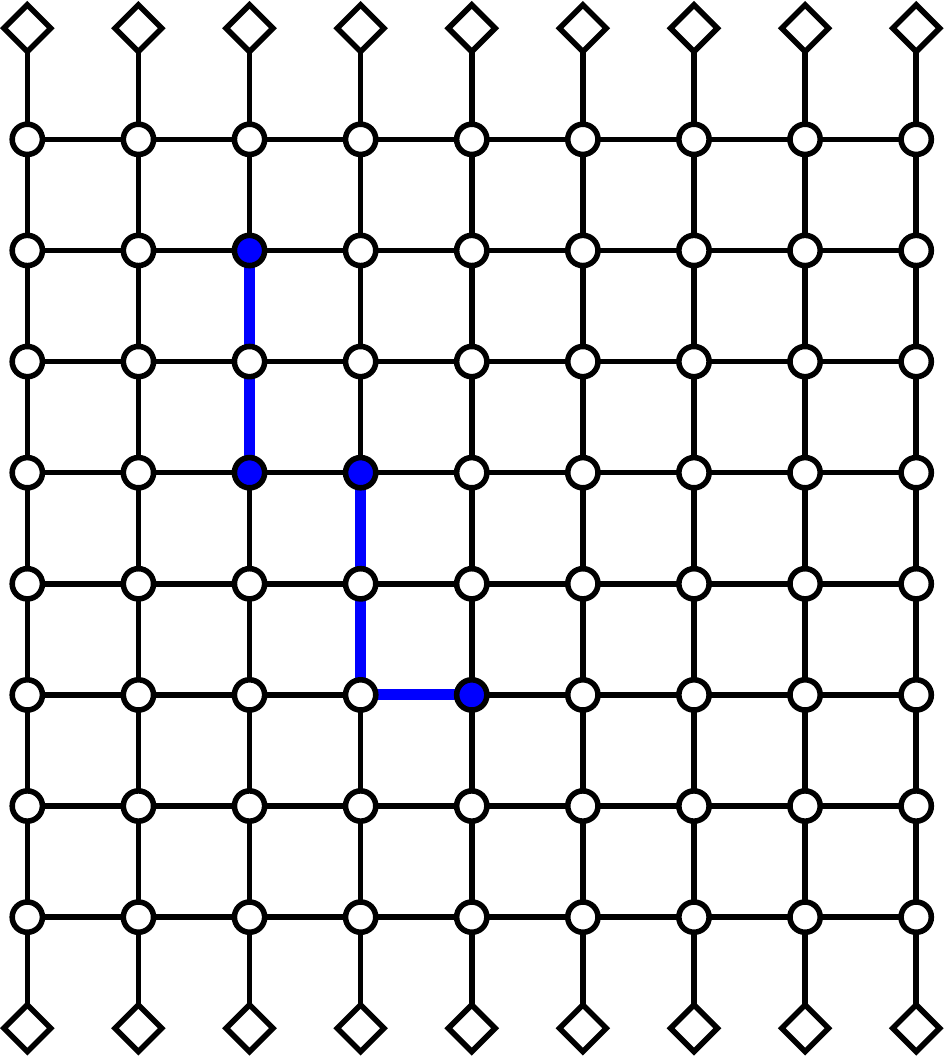}
\caption{Implementation of the Min-Weight Perfect Matching decoder for the error presented in Fig.~\ref{fig:SCPrimal}. The decoder optimizes to find smaller weight corrections than the Greedy decoder by searching for pairs of higher weight.}
\label{fig:SCDualMWPM}
\end{figure}

In topological codes, error syndromes are registered as excitations of stabilizers on the topological surface in question. That is, given an error chain composed of a particular Pauli operator, the resulting syndromes whose measurement results will change will be the end points of the error chain. We denote such changes in the expectation value of the syndromes as excitations. Treating flipped syndromes as excitations is rooted in the fact that the endpoints of these error chains behave like anyonic particles, with associated braiding statistics depending on the type of error. As such, any decoder will need a notion of the topology at hand. The code distance will refer to the weight of the logical Pauli operator with smallest support. between different excitations will allow to minimize the probability of introducing a logical error upon decoding. In Fig.~\ref{fig:SCPrimal}, we provide an example of a weight-4 $X$~error on a distance~9 surface code patch. Given that the error is of low weight, proper decoding should lead to its correction. The inputs to the decoding algorithm are most simply described in the dual lattice picture, as seen in Figs.~\ref{fig:SCDualGreedy}--\ref{fig:SCDualMWPM}, where lattice sites correspond to syndrome locations, and edges correspond to shared qubits between pairs of stabilizers. 

The input to the decoding algorithm is the locations of the excitations, or flagged syndromes, and the corresponding distance between pairs of excitations. The distance between excitations is given by the minimal number of qubits connecting a chain with the excitations at the endpoints. The simplest form of decoder - the Greedy algorithm - leads to poor performance. Given a look-up table of all the excitation pairs and their corresponding distances, the Greedy decoding algorithm finds the smallest weight pair, and assigns a correction along the connecting chain. Then, these two excitations are removed from the table, and the next-smallest pair is found, and so-forth until all pairs of excitations have been corrected. In the example provided in Fig.~\ref{fig:SCPrimal}, the two neighboring excitations are first matched up together (since they are distance~1 apart), and then the remaining excitations are matched up through the boundary (as this forms the minimal-weight error between these excitations). Unfortunately, when combined with the initial errors in the code, the remaining error will now form a logical error since the Greedy decoder accidentally paired syndromes that were quite delocalized.

A more accurate form of correction can be achieved using an algorithm called Min-Weight Perfect Matching~(MWPM). The MWPM algorithm is based of Edmonds' blossom algorithm~\cite{Edmonds.1965}, where excitations are matched together by considering growing sets of matchings of different weights. Moreover, the algorithm is efficient in the number of input pairs, scaling roughly as~$N^6$, where $N$ is the total number of excitations\footnote{the blossom algorithm scales as $P^3$, where $P$ is the total number of pairs. Since there are $N$ excitations, there are on the order of~$N^2$ pairs of excitations.}. In the example shown in Fig.~\ref{fig:SCDualMWPM}, the MWPM ~algorithm matches together two pairs of excitations that are of distance~$2$ and~$3$. Therefore, the total distance of the corresponding pairs is~$5$, as opposed to in the case of the Greedy decoder, where the total correction was of weight~$1+5 = 6$. The MWPM~algorithm aims to minimize over this sum.

\subsection{Qudit Codes}
Qubit stabilizer codes can be generalized to the case of \emph{qudits} through the introduction of generalized Pauli operators, the Heisenberg-Weyl operators. A qudit resides in a $D$-dimensional Hilbert space, and the generalized Pauli operators take the form:
\begin{equation}
X = \sum_{j=0}^{D-1} \kb{j \oplus 1}{j}, \qquad Z = \sum_{j=0}^{D-1} \omega^j\kb{j}{j},
\end{equation}
where addition is taken mod~$D$ and $\omega:= e^{2\pi i/D}$ is the $D$-th complex root of unity. Intuitively, the $X$~and~$Z$ operators generalize the regular Pauli operators, and while no longer being self-inverse, do satisfy~$X^D = Z^D = I$. Stabilizer codes will again be defined as the ``+1''~eigenspace of the generalized Pauli operators. Importantly, the operators no longer just simply commute or anti-commute, but rather have a generalized form of commutation relationship, $X^k Z^l = \omega^{-kl} Z^l X^k$. Therefore, when generalizing topological CSS codes to the setting of qudits, it will no longer be sufficient for $X$ and $Z$~type stabilizers to just overlap at an even number of locations, as the phases will no longer automatically cancel out. In order to overcome this restriction, particular orderings for each of the operators are chosen in order to satisfy commutativity of the stabilizers~\cite{Kitaev20032}. We will expand upon this notion further in upcoming sections. One distinct advantage of qudit codes is that due to the enlarged local Hilbert spaces, stabilizer measurements can return a much larger set of possibilities. As such, increased information about chains of errors can be detected by neighboring plaquettes, since a weight-2 error on a given stabilizer will not necessarily lead to a trivial syndrome measurement. This phenomenon has been shown to produce higher thresholds for qudit codes, in comparison with similar qubit codes, when decoding takes this additional information into account~\cite{watson2015fast}. 

Emphasizing this increased difficulty in decoding, the min-weight perfect matching decoding technique described in the previous section translates to hyper-graph matching in the case of qudit decoding on the surface code, a computationally inefficient task. As such, modified decoding techniques have to be used in the case of qudit codes, and these are typically based on a form of renormalization~\cite{watson2015fast}. The process involves iteratively increasing the scale, at each scale identifying maximally connected disjoint clusters, and if they are neutral (the sum of excitations~$= 0 \mod D$), annihilating them through charge-transport, as described in more detail in a later section.

Just as in the case of qubit codes, topological qudit codes exhibit thresholds due to the percolation of errors. That is, given a particular fixed topological code and decoder, for independent errors at a rate below a certain threshold, the code will be corrected with high accuracy, reducing the logical error rate of the state with respect to the physical error rate. However, once the statistical fluctuations of errors become too great, larger error regions will form, leading to high likelihood that such large error regions will be connected to form a logical error that spans non-locally across the system~\cite{dennis2002topological}. The rate at which the system undergoes these phase transitions is denoted by the threshold for the system, given a particular error model. The threshold rate can greatly vary depending on the error model of the system, commonly spanning one or more orders of magnitude when comparing independent memory errors from circuit level noise (that is noise on the individual circuit components that are used to perform the error correction and detection).

\section{Decoding Qudit Color Codes\label{sct:gsp}}

In this work, we use clustering techniques to decode the qudit color code. We will call this the GCC Decoder, for Generalized Color Clustering. In this description, we will focus on its application to the 6-6-6 code, but it can be just as easily applied to the 4-8-8 or 4-6-12 codes. We would like to point out that the goal of our decoding algorithm will be to find a correction process that will, along with the original error, return the state of the system to the codespace. The resulting correction therefore must combine with the error to result in no stabilizers being violated, or in the language of charge excitations, to cancel out all charge excitations through transport and fusion. In this section, we will discuss moving charges around by applying Pauli gates. These gates do not necessarily have to be applied physically, but rather can be tracked in software throughout further computations and it is this setting we have in mind when discussing these algorithms.

\begin{figure}
\includegraphics[width=0.12\textwidth]{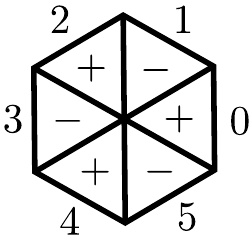}
\caption{Single 6-6-6 code plaquette in the dual picture, with measurement qudits at vertices and data qudits at the center of each triangle. We define a counterclockwise orientation and an ordering of the data qudits with respect to the measurement qudit.}
\label{fig:Orientation}
\end{figure}

For plaquettes in qudit-valued topological codes to be stabilizers, we need to define an orientation on the code. For example, we can label the six data qudits surrounding each measurement qudit with numbers $0$ through $5$, increasing counterclockwise starting from the positive $x$ axis. With this ordering in hand, we can assign positive signs to data qudits with even parity, and negative to data qudits with odd parity. This means that when we act with SUM gates on a measurement qudit - data qudit pair, we would add the charge of qudit $0$, but subtract the value of qudit $1$ from the total charge of the target measurement qudit. Fig.~\ref{fig:Orientation} gives an example of such an orientation, represented in the dual picture.

In order for this to be a commuting stabilizer code, as explained thoroughly in the literature~\cite{brellnonabelian}, we must choose one \textit{privileged} color (usually red) to have the opposite sign convention as the other two, associating odd parity with positive. This is important to guarantee the commutativity of generalized $X$ and $Z$ plaquettes in the qudit code and can be tracked in software.

This orientation allows us to generate a valid and meaningful syndrome from our data errors on the color code. For example, if during our code cycle only a single data qudit error of magnitude $k$ acted on our code, this would trigger the three surrounding measurement qudits in the following way: For B, R, and G, the data qudit is at (odd) positions $1$, $3$ and $5$, so during the code cycle, each of these triggers $-k$, where $-k$ is the additive inverse of $k$ modulo qudit dimension $D$. In general, a given data qudit will be at positions of the same parity for each of the neighboring measure qudits.

\begin{figure}
\includegraphics[width=0.12\textwidth]{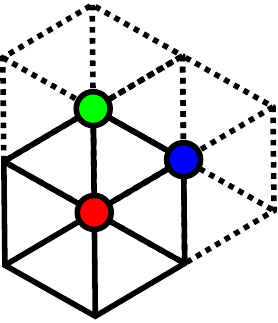}
\caption{A given data qudit is at different position in the ordering for each of the surrounding measurement qudits.}
\label{fig:DifferentPositions}
\end{figure}

Inverting this process tells us how to transport charge along the lattice. The simultaneous transport of charge~$k$ from R to the two other measurement qudits in the triangle is accomplished by changing the value of the data qudit enclosed by the triangle (which can be performed with post-processing software). In particular, we must subtract the signed charge $k$ (modulo~$D$) from the data's initial charge. The sign, as above, is determined by the parity of the data qudit with respect to R. In this process, both B and G pick up a charge of~$-k$ regardless of data qudit sign.

\begin{figure}[h!]
\includegraphics[width=0.15\textwidth]{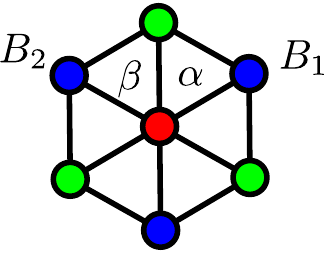}
\caption{Red plaquette with data qudits $\alpha$ and $\beta$ at positions $1$ and $2$. We can transport charge from $B_1$ to $B_2$ by modifying the values of $\alpha$ and $\beta$ in the software.}
\label{fig:TransportCharge}
\end{figure}

This leads us to the following key observation: by composing such transport processes, we can transfer charge between qudits of the same color without affecting measurement qudits of complementary types. When clustering in the surface code, we were able to annihilate a neutral cluster through controlled charge transport. This same-color transport allows us to perform a similar procedure for qudit color codes.

\begin{center}
    \includegraphics[width=0.45\textwidth]{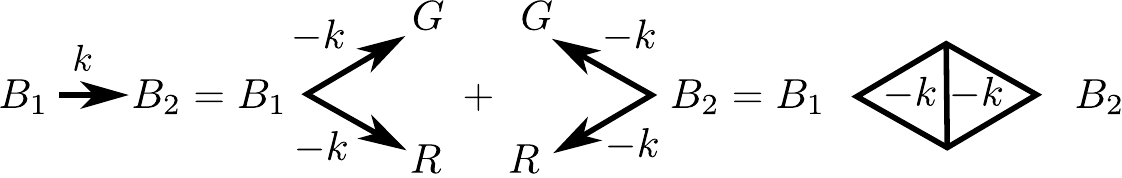}
    \captionof{figure}{Transporting charge $k$ from $B_1$ to $B_2$ is equivalent to the combination of transporting $k$ from $B_1$ to R and G, and transporting $-k$ from $B_2$ to R and G.}
\end{center}

Here, adding $k$ to qudits $\alpha$ and $\beta$ transports charge $k$ from $B_1$ to $B_2$. In this way, we can transport charge across greater distances on the lattice by composing these transportations.

\begin{figure}[h!]
\includegraphics[width=0.22\textwidth]{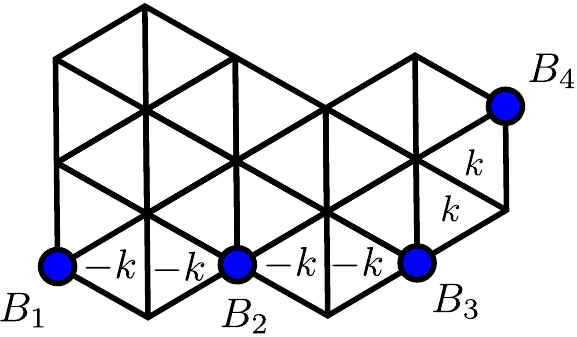}
\caption{Illustration of charge transport from $B_1$ to $B_4$. Note that between $B_3$ and $B_4$ we add charge $k$ to both data qudits because the parity is odd according to our specified orientation.}
\label{fig:LongTransport}
\end{figure}

\subsection{GCC Decoder}

At this point we can introduce our decoder. Similarly to the surface code case, we will have a notion of \textit{boundary-neutral}, however for the color code this will be slightly more nuanced. In the surface code, we called a cluster boundary-neutral if the cluster contained \textit{any} boundary element. We proceeded to annihilate all boundary clusters before iterating the scale. For the color code, even if a cluster contains a boundary element, we can not necessarily annihilate the entire cluster. Therefore, we will refer to a cluster as boundary-neutral only when we can annihilate the cluster through internal charge transport and transport with the included boundaries.

The decoding proceeds as follows: We define our primitive unit of distance as the length of an edge on the dual lattice $l$. Starting with $i=1$ (the scale factor), we connect all measurement qudits in the syndrome separated by distance $\leq l\times i$, and identify all  maximally disjoint clusters. Within each cluster, we calculate the average position, and then transport the charge from all elements in the cluster toward the measurement qudits closest to that central point\footnote{We also tried checking the net charge of a cluster first and as in the surface code case, only transporting charge in neutral clusters. However, empirically our collapse method, which preserves distance between the centers of clusters, gave better results.}, making corresponding data qudit changes in the software as we go. If the cluster is neutral, it will have been completely annihilated. Otherwise, we are left with at most two qudits (of distinct colors) at the center of the cluster. Boundary-neutral clusters are then annihilated, and the scale is iterated to the next integer value. the procedure is repeated until the syndrome is empty.

To determine if a particular decoding was successful (under a code capacity error model), it is sufficient to check that the original error plus correction commutes with a single logical operator (any of the three boundaries of the triangular code). An explicit example of a successful GCC decoding is given in Appendix \ref{app:GCC}.

\section{Threshold Estimates for Qudit Color Codes\label{sct:results}}
The performance of the different codes and decoders are tested in the memory bit-flip error model, for a variety of qudit dimensions. We only study the case of bit-flip noise as the code is a CSS code, and our decoding techniques would address errors of bit and phase type independently. Considering one type of noise is standard in estimating the performance of CSS codes under code~capacity noise and will give us a natural comparison point to previous studies of color~code decoders~\cite{PhysRevA.89.012317}. The memory model assumes noisy qudit memories, with perfect syndrome extraction and correction (according to the chosen decoder), and is generally considered to be a good representative of the performance of families of error correcting codes. That is, good performance under this error model will typically translate into good performance under gate noise when using fault-tolerant measurement circuits, although the threshold will typically decrease by at least an order of magnitude. Unless otherwise specified, each logical error probability data point represents 30000 trials. 

Explicitly, the error model we are studying here, in the case of qubits, is the following applied to each data qubit of the code:
\begin{equation}
\mathcal{E}(\rho) = (1-p) \rho + p  X \rho X.
\end{equation}
Namely, the error model consists of random bit-flip noise being applied with probability~$p$.

For qudits, each qudit in the code can be affected by $D-1$ unique non-trivial integer multiples of each generalized bit-flip~$X$ operator. As a result, the error model generalizes to: 

\begin{align}
\mathcal{E}(\rho) = (1-p) \rho &+ \dfrac{p}{D-1}\sum_{j=1}^{D-1}X^{j} \rho X^{-j}.
\end{align}

To put our decoding results in context, we first present results for the surface code using min-weight perfect matching~(MWPM) and clustering techniques, and then present qubit color code results using the surface projection decoder of Delfosse~\cite{PhysRevA.89.012317}, which relies on min weight perfect matching having projected the initial color code syndrome information onto multiple copies of the surface code.

\subsection{Surface Code Threshold Estimates}
The first set of thresholds we present are for two different decoders in the surface code. While the min-weight perfect matching decoder provides an optimal solution for the presence of excitations in the stabilizer dual graph caused by errors, its generalization to higher dimensional systems does not scale well, as previously discussed. As such, in parallel we present the clustering algorithm which will be used at higher dimensions, and will take a hit in terms of threshold rate in the case of qubits~$(D=2)$. Our simulation of the MWPM decoder aligns with previous theory and simulations, converging to a threshold value of~$0.103$.

\begin{figure}[h!]
\includegraphics[width=0.4\textwidth]{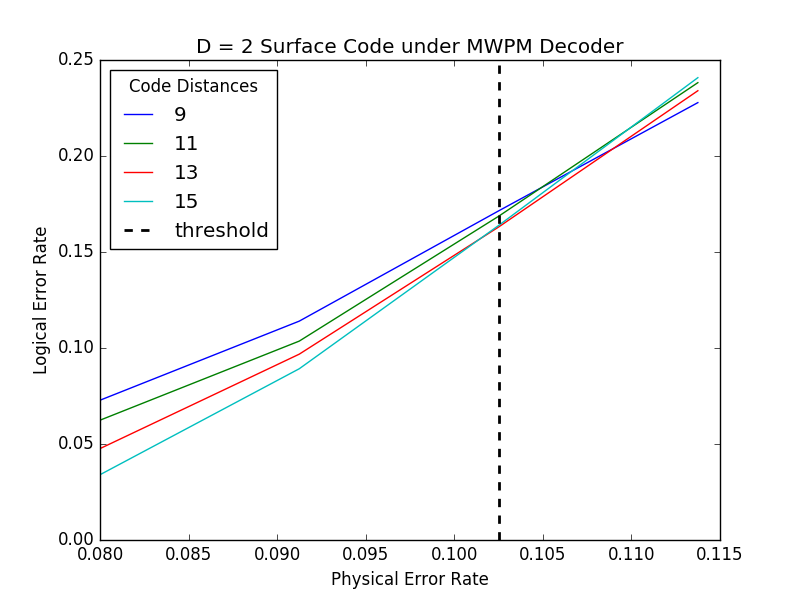}
\caption{Logical error rates for Kitaev surface code under min weight perfect matching decoder, which identifies maximally likely error chains. We have identified the threshold as the crossing of the distance 13 and 15 codes, at $p_{thresh} = 0.103$. Each data point corresponds to only 5000 trials, and trials were taken at fewer probabilities due to the relatively large computational cost of min weight matching as compared to renormalization techniques.}
\label{fig:mwpm}
\end{figure}

\begin{figure}[h!]
\includegraphics[width=0.4\textwidth]{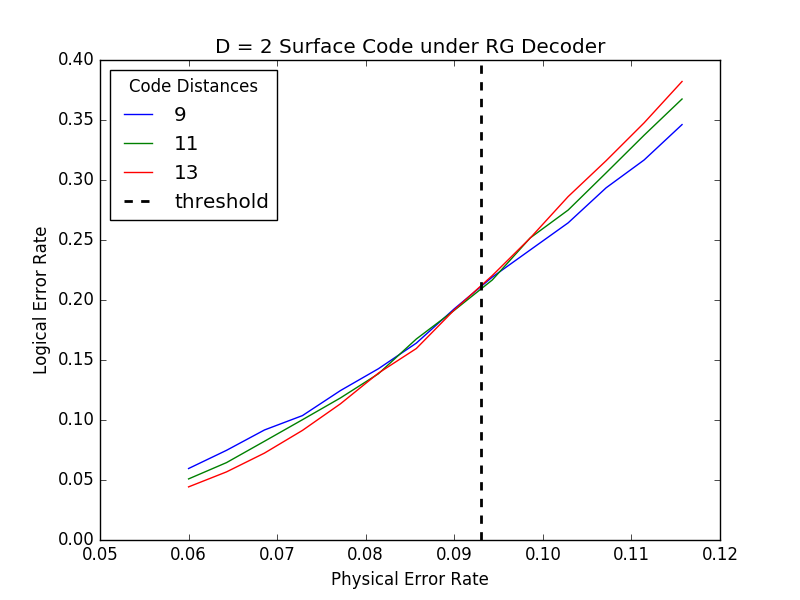}
\caption{Logical error rates for Kitaev surface code under hard decision renormalization group decoder. As expected, in this qubit case, clustering slightly under-performs compared to the minimum weight perfect matching decoder, giving a threshold at $p_{thresh} = 0.093$}
\label{fig:rg2}
\end{figure}

\begin{figure}[h!]
\includegraphics[width=0.4\textwidth]{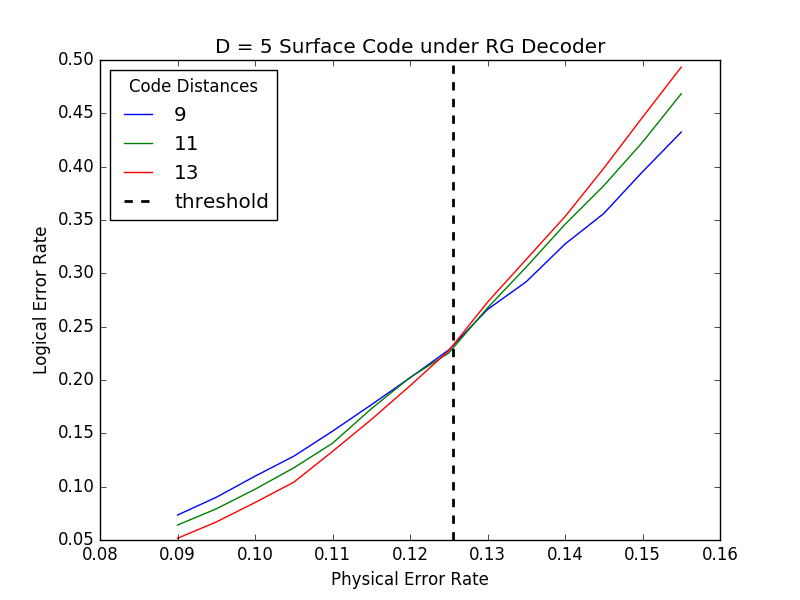}
\caption{Kitaev surface code under hard decision renormalization group decoder for qudit D=5. $p_{thresh} =0.1255$}
\label{fig:rg5}
\end{figure}

\begin{figure}[h!]
\includegraphics[width=0.4\textwidth]{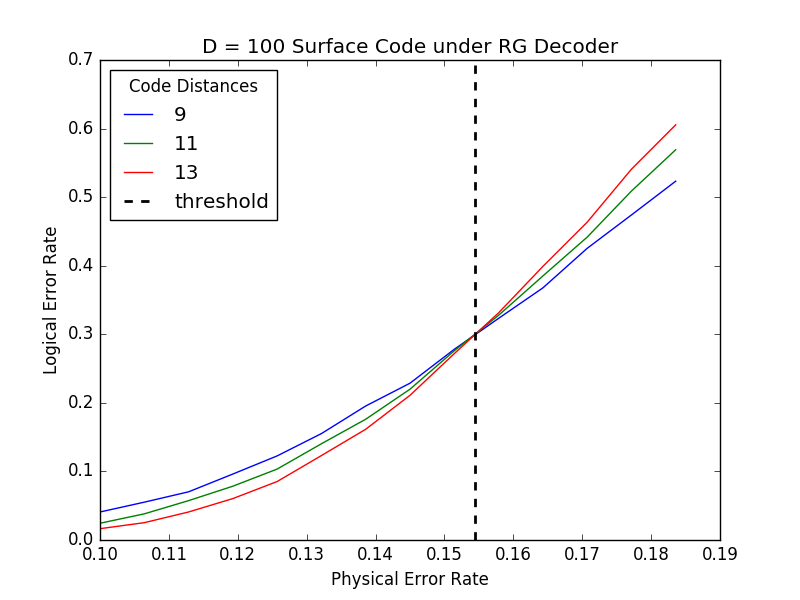}
\caption{Kitaev surface code under hard decision renormalization group decoder for qudit D=100. At $p_{thresh}= 0.1545$, we have almost reached a plateau.}
\label{fig:rg100}
\end{figure}

\begin{figure}[h!]
\includegraphics[width=0.4\textwidth]{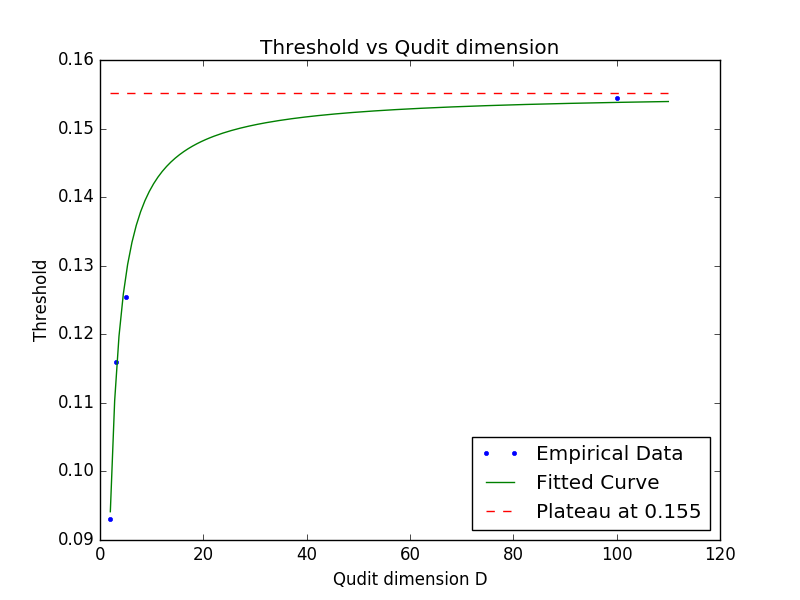}
\caption{The threshold increases with qudit dimension until saturating around $p_{thresh} = 0.155$. Curve fitted to function $ T(x) = T_{plateau} - \frac{\alpha}{\beta - D}$, where $D$ is the qudit dimension, $T$ is the threshold, and $T_{plateau}$ is the threshold as $D \rightarrow \infty$.}
\label{fig:rg_thresh}
\end{figure}

In contrast, the clustering algorithm based on renormalization yields a threshold result of~$0.093$ in the case of qubits. However, this decoding algorithm can then be generalized to higher-dimensional qudit systems, as outlined in the previous sections. The shift in the threshold value obtained from the crossing point of the logical error rate at various distances can be explicitly seen as the qudit dimension increases, see Figs.~\ref{fig:rg5}--\ref{fig:rg100}. The threshold value increases monotonically as the qudit dimension rises, saturating close to a threshold value of~$0.155$ as shown in Fig.~\ref{fig:rg_thresh}.

\subsection{Color Code Threshold Estimates}
Before describing decoding methods for qubit and qudit color codes, we caution against direct comparison of specific threshold values between the surface codes discussed above and 6-6-6~ color codes, as they have different underlying geometries. The Delfosse Surface Projection~(DSP) method for decoding the qubit color code is based on the idea of projecting two of the three colors onto a surface code, and then using a surface code decoder to obtain a set of different correction paths. Unifying the correction paths from these different projections determines a valid correction for the color code, at the expense of reducing the threshold value~\cite{PhysRevA.89.012317}. This decoding method achieves the highest known decoding rate in terms of code capacity noise for the color code. There are two potential methods then to generalize such a  decoding technique to higher dimensional qudits, either by using similar projections and a generalized decoder at the surface code level, such as the renormalization decoder analyzed in the surface code, or to generalize the clustering algorithm directly. We chose the latter, adapting the renormalization group decoder into a General Color Clustering~(GCC) decoder. 

\begin{figure}[h!]
\includegraphics[width=0.4\textwidth]{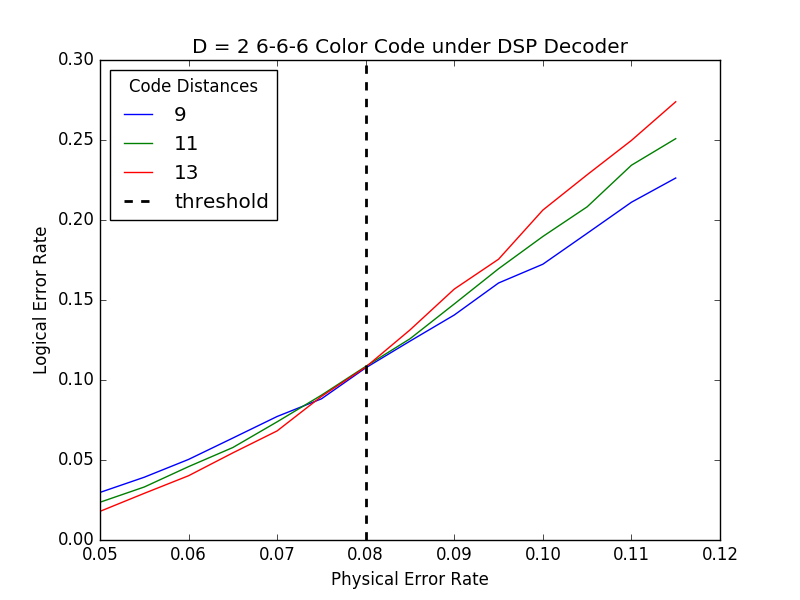}
\caption{Threshold estimates using the Surface Projection decoder, which employs min weight perfect matching on each of the code's shrunk lattices. $p_{thresh} = 0.080$}
\label{fig:dsp}
\end{figure}

\begin{figure}[h!]
\includegraphics[width=0.4\textwidth]{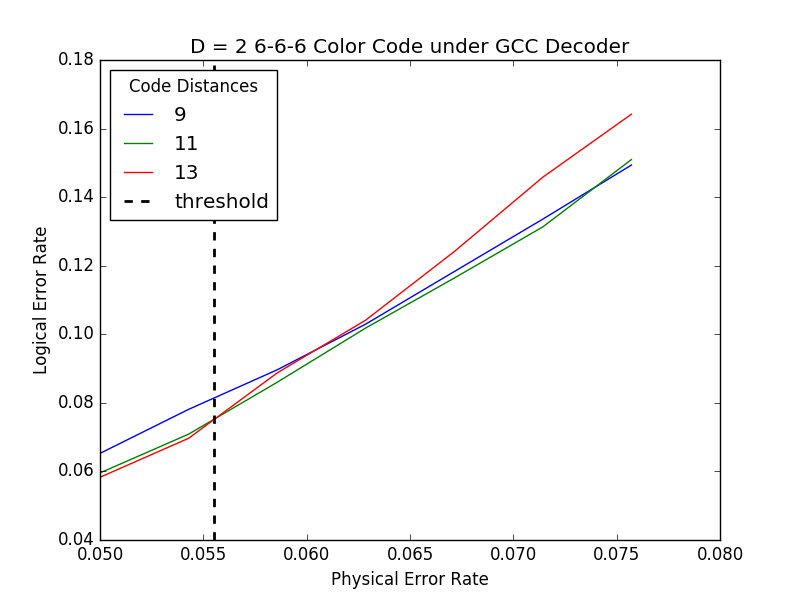}
\caption{6-6-6 color code under general color clustering decoder for qubit D=2. $p_{thresh} = 0.056$}
\label{fig:gcc2}
\end{figure}

\begin{figure}[h!]
\includegraphics[width=0.4\textwidth]{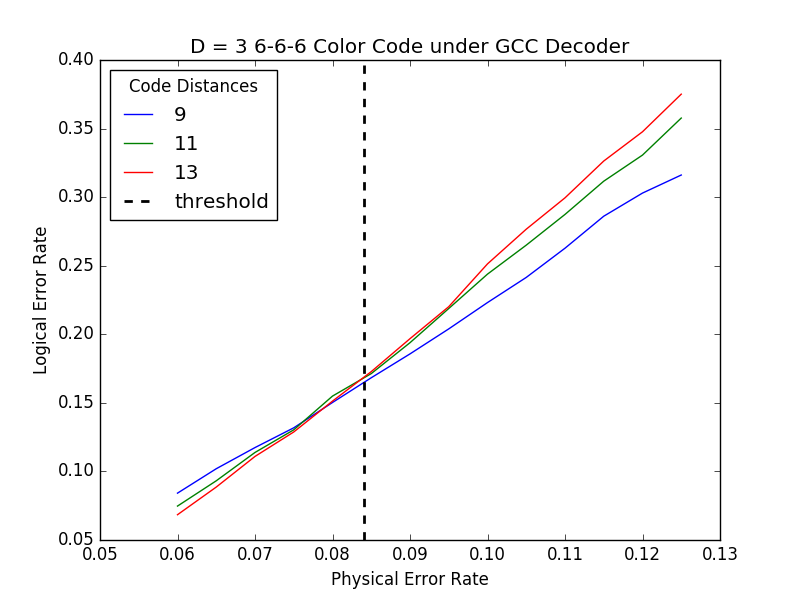}
\caption{6-6-6 color code under general color clustering decoder for qudit D=3. $p_{thresh} = 0.084$}
\label{fig:gcc3}
\end{figure}

\begin{figure}[h!]
\includegraphics[width=0.4\textwidth]{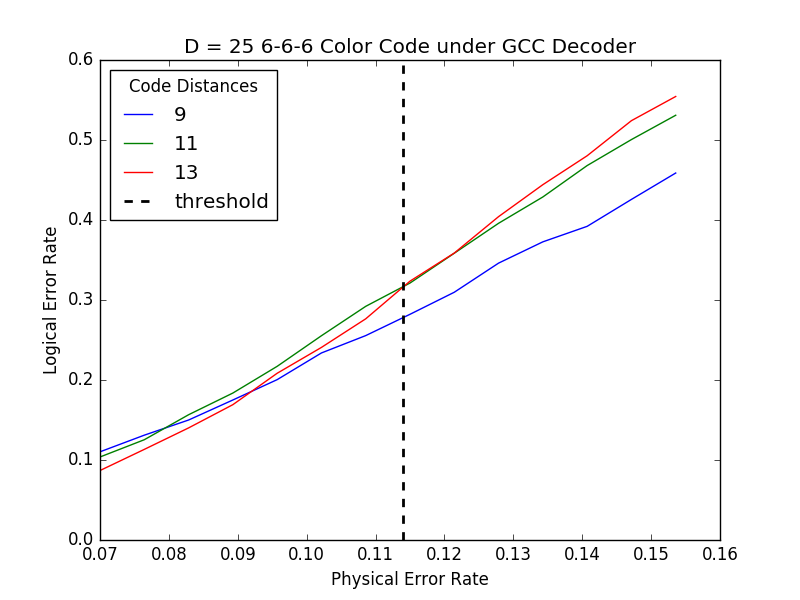}
\caption{6-6-6 color code under general color clustering decoder for qudit D=25. $p_{thresh} = 0.115$}
\label{fig:gcc25}
\end{figure}

\begin{figure}[h!]
\includegraphics[width=0.4\textwidth]{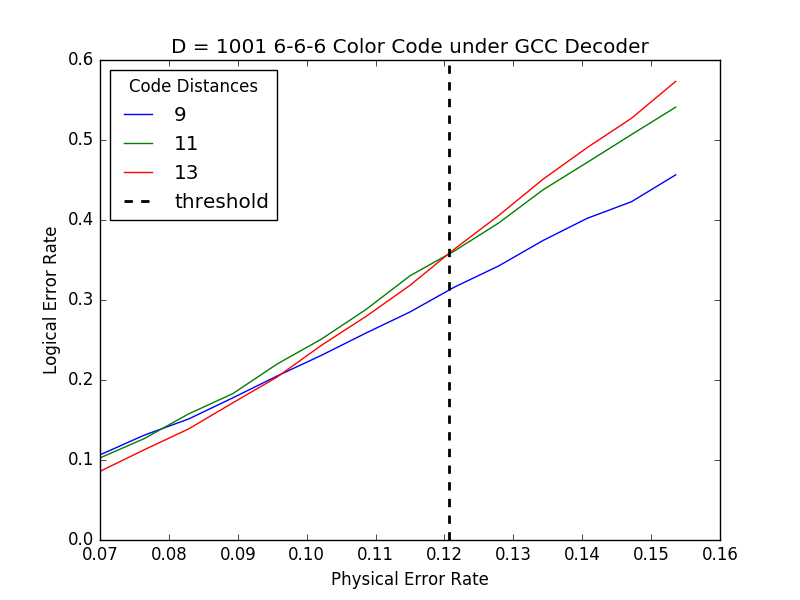}
\caption{6-6-6 color code under general color clustering decoder for qudit D=1001. $p_{thresh} = 0.1207$}
\label{fig:gcc1001}
\end{figure}

The results for the color code decoders are presented in Figs.~\ref{fig:dsp}--\ref{fig:gcc1001}. The implementation of the DSP decoder realizes a threshold rate of~$0.080$ for code capacity noise in the qubit model, slightly below the result obtained in theory and simulation. This threshold, which applies only to \textit{our} particular implementation differs from Delfosse's result of ~$0.087$. Yet even the threshold for our DSP implementation exceeds the GCC decoder in the case of qudits. Moreover, in the limit of large qudit dimension the thresholds obtained from the GCC decoder surpass that of the ideal qubit color code decoding.

\begin{figure}[h!]
\includegraphics[width=0.4\textwidth]{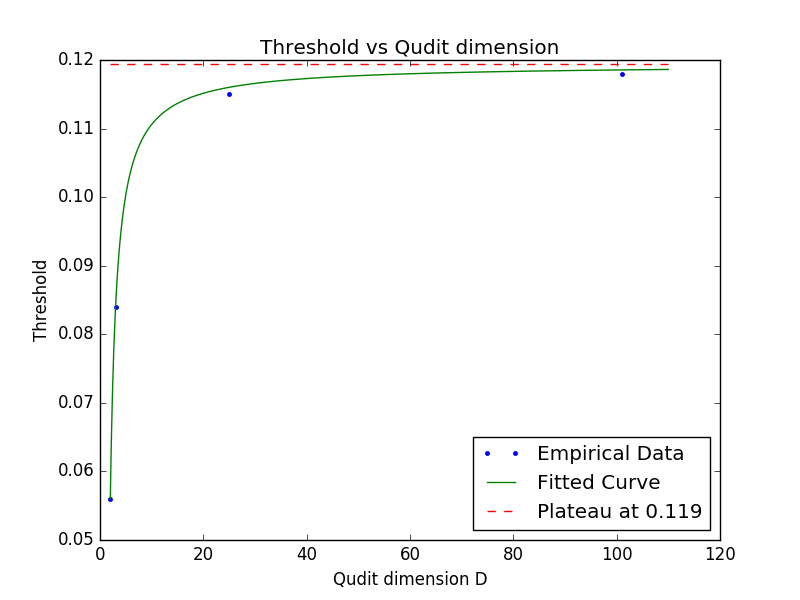}
\caption{As in the surface code case, the threshold increases with qudit dimension until saturating. The plateau occurs at $p_{thresh} = 0.119$. Curve fitted to same form as in Fig. \ref{fig:rg_thresh}, including the data point at $D=1001$ not shown in this plot.}
\label{fig:gcc_thresh}
\end{figure}

The Surface Projection decoder for qubit color codes works by splitting a color code into three shrunk lattices, and performing MWPM on each lattice. Just like the Hard-Decision renormalization group algorithm for surface codes, our clustering algorithm approximates a hard problem by breaking up a global error syndrome into smaller, localized correction clusters. In analogy to the surface code case, then, we would expect our GCC decoder to attain a slightly lower qubit threshold than the DSP decoder. In addition, we should see threshold increasing with qudit dimension. In fact, we see just this: the clustering decoder exhibits a lower threshold for qubits, attaining a rate of ~$0.056$, which is an even more substantial similar drop from ideal threshold value than in the surface code case. However, in generalizing this decoding technique to higher dimensional systems, the threshold rate increases to~$0.084$~$(D=3)$, and to~$0.115$~$(D=25)$, eventually plateauing at $0.119$, see Figs.~\ref{fig:gcc2}--\ref{fig:gcc_thresh}.

\begin{table}
\begin{tabular}{|c|c|c|c|}
\hline
 Code & Decoder & $D = 2$ & $D \rightarrow \infty$ \\ 
 \hline
 \hline
 Surface Code & MWPM & 0.103 & \\  
 \hline
  & HDRG & 0.093 & 0.155\\ 
  \hline
  \hline
 $(6.6.6)$ Color Code & DSP & 0.080 & \\  
 \hline
  & GCC & 0.056 & 0.119\\ 
  \hline
\end{tabular}
\label{tb:Thresholds}
\caption{Code capacity threshold results for the implemented simulations of the surface and (6.6.6)~color codes. In the case of the surface code, the optimal Min-Weight Perfect Matching~(MWPM) is compared to the Hard-Decision renormalization group~(HDRG) decoder for qubits~$(D=2)$, and the latter is then generalized to higher qudit dimension. Similarly, for the color code, the Delfosse Surface Projection method~(DSP) is compared to the General Color Clustering~(GCC) decoders and further generalized.}
\end{table}

Unlike in the surface code, in the color code of code distance~$7$, every internal measurement qudit is connected to at least one boundary from the first clustering iteration onward. Because of this, our decoder performs poorly. These \textit{small-size} effects become less important for code distance~$9$, and vanish almost entirely for larger codes. In order to properly reflect these small-size effects, we do not consider this curve when establishing the threshold.  However due to length of the classical simulation we were only able to simulate lattice sizes up to code distance~13. We used the crossing point of the code distance 11 and 13 as a point of reference for the threshold value and believe that further simulations at higher values would strengthen the assumption that the threshold value is at this point.

\section{Conclusion\label{sct:conclusion}}
In this work, we have introduced a new decoding algorithm for qudit color codes to study the behavior of the fault-tolerance threshold as qudit dimensionality is increased. Since algorithms that use Min-Weight Perfect Matching as a primitive, such as the Delfosse Surface Projection method~\cite{PhysRevA.89.012317}, do not scale well with increased dimension, a new scheme based on renormalization group techniques is proposed called General Color Clustering~(GCC). This mirrors similar work completed by Watson~\textit{et~al.} in the case of surface codes~\cite{watson2015fast}, and similar conclusions are drawn. Namely, the memory threshold drops in the case of qubits from ~$0.080$ in the case of the surface projection method used in our simulations to ~$0.056$ for the case of~GCC. However, as qudit dimension is increased, the threshold rate rises to a saturation point of near double that of the qubit case, that is~$0.119$. We expect that the decoder would exhibit a similar behavior when generalizing the noise to include measurement errors, phenomenological noise, however leave this study for future works. 

It should be noted that while the threshold value for qudit codes exceed that of the qubit base, a direct comparison may be misleading. Namely, while the probability of introducing an error is the same, the probability of a given error type is reduced in the qudit case since there are an increased number of possible errors due to the growth in system size. One would expect any experimental implementation of qudit codes to be hampered by the increased dimensionality of the system size in terms of potential error leakage, as there are increased degrees of freedom that can be coupled to the environment. Whether such noise can be reduced to sufficiently small levels in order to take advantage of the properties that qudit codes provide remains an interesting question for experimental implementations.

For the qudit surface code, the effect of adding initialization steps to their clustering decoder to account for path degeneracies has additionally been studied. It would be interesting to perform a similar analysis with our qudit color code decoder, as in the 4-8-8 and 4-6-12~codes certain error chains are exponentially suppressed. 

Going forward, we would also like to generalize our clustering algorithm to 3D qudit color codes, and to explore possible uses of clustering in qudit gauge color codes.

\begin{acknowledgments} 
The authors thank Michele Mosca for fruitful discussions and comments regarding the manuscript, and the IQC technical staff for providing access to the Heavylift computing clusters at the IQC. We also thank Earl Campbell and Benjamin Brown for useful comments. Jacob Marks acknowledges  support from a Yale Richter Fellowship and Undergraduate Research Award from the University of Waterloo. Tomas Jochym-O'Connor acknowledges support from NSERC through the Vanier CGS as well from the Burke Institute through the Sherman Fairchild Fellowship. Vlad Gheorghiu acknowledges support from NSERC and CIFAR. IQC is supported in part by the Government of Canada and the Province of Ontario.
\end{acknowledgments}

%

\pagebreak
\appendix

\section{General color clustering decoder}
\label{app:GCC}
\begin{figure}[h!]
\includegraphics[width=0.4\textwidth]{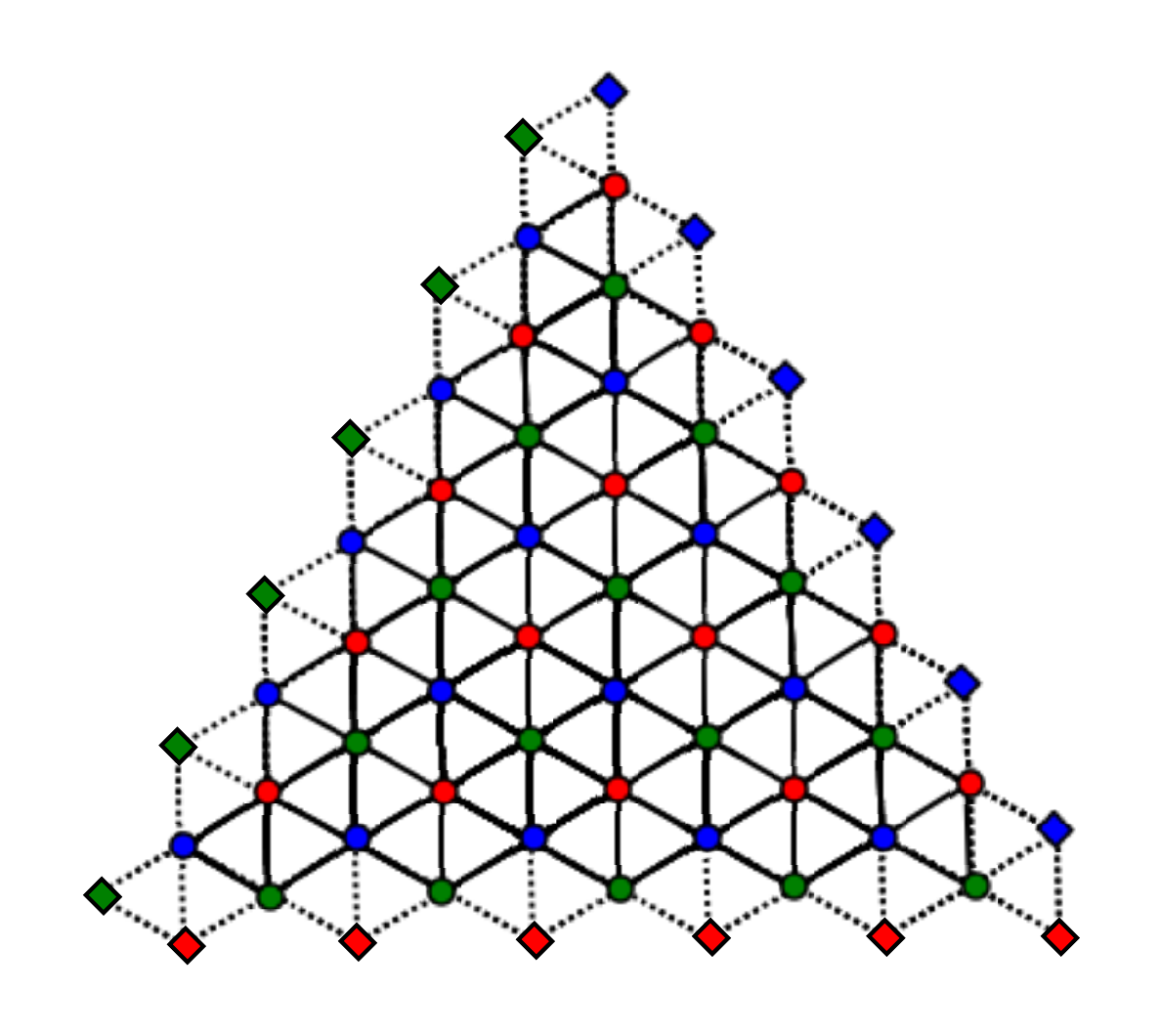} 
\caption{The dual lattice for the 6-6-6~color code with distance~11. The vertices with circles represent stabilizers of a given color, while those with diamonds represent the appropriate boundary. Data qudits reside at plaquettes, therefore each stabilizer measures 6~data qudits (or less at the boundary).}
\label{fig:CCDual}
\end{figure}

In this Appendix, we provide some additional detail on the general color clustering~(GCC) used to decode qudit color codes. As mentioned in the main text, the decoding algorithm takes as inputs the qudit stabilizers that have been violated, and uses techniques from renormalization group decoders to cluster sets of corrections. Fig.~\ref{fig:CCDual} provides a graphical representation of the dual lattice of the 6-6-6~color code with distance~11, where qudit stabilizers measure the data errors according to the orientation described in Section~\ref{sct:gsp}.

\begin{figure}[h!]
\includegraphics[width=0.4\textwidth]{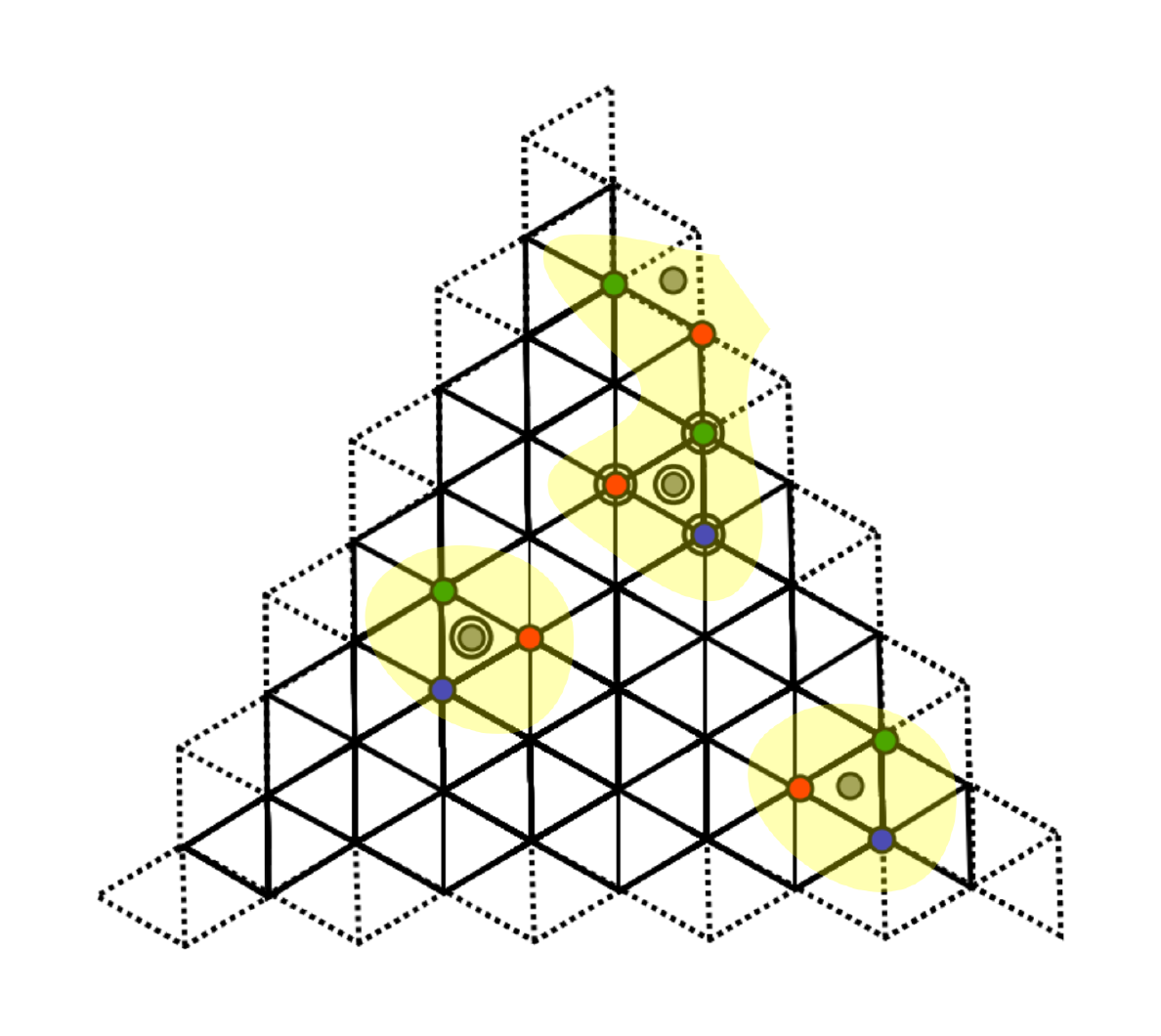} 
\caption{Example of a set of errors for the qudit color code of dimension~3. Errors are given in grey, and are of a single type, say~$X$. The errors of charge~1, that is~$X^1$, are represented by circles with a single contour line. Errors of charge~2, that is~$X^2$, are represented by double contour lines. Only violated stabilizers are shown in this graph, and their charge is represented as in the case of the qudit errors.}
\label{fig:GCCStep0}
\end{figure}

A set of qudit errors triggers the appropriate set of measurement flags which will serve as input to the decoder. Fig.~\ref{fig:GCCStep0} provides an example of the stabilizer measurements observed in a qudit color code with dimension~3. Since the GCC~decoder acts as a CSS~decoder, addressing errors of $X^k$ and $Z^k$~type separately, we will assume for simplicity that all of the errors are of the same type, say~$X$. The qudit errors are represented by grey circles, where a single line contour represents a ~$X^1$ error, while a double contour represents a $X^2 \cong X^{-1}$~error. Only the violated stabilizers are shown in the figure, and they register the appropriate charge according to the orientation of the measured errors. In the example, the error in the bottom right is of type~$X^1$, and since it is in the positive orientation of the neighboring stabilizers, they register violations corresponding to charge~1. However, in the error of charge~2 on the left, the neighboring stabilizers register the error as of charge~(-2), therefore equivalent to charge~1.

The clustering algorithm takes all input stabilizers and their associated graph, and clusters all violated stabilizers that are within distance 1 of each other (that is, each cluster must consist of violated stabilizers that are at least of distance~2 from any other cluster). All charges within the cluster are then brought together to three geometrically central nodes, that is one for each color. This is done through charge transport. In our algorithm, we reduce the three central charges to at most two central charges by canceling out at least one of the charges through charge identities between the three colors. If a given cluster is charge neutral, then all charges are removed and the correction is complete (for that cluster). Any remaining charge is then iteratively fed back into the same algorithm, where clusters are now formed by joining charges that are distance~2 apart. The process continues until all charges are removed. In Fig.~\ref{fig:GCCStep0}, the yellow regions represent the different clusters. 

\begin{figure}[h!]
\includegraphics[width=0.4\textwidth]{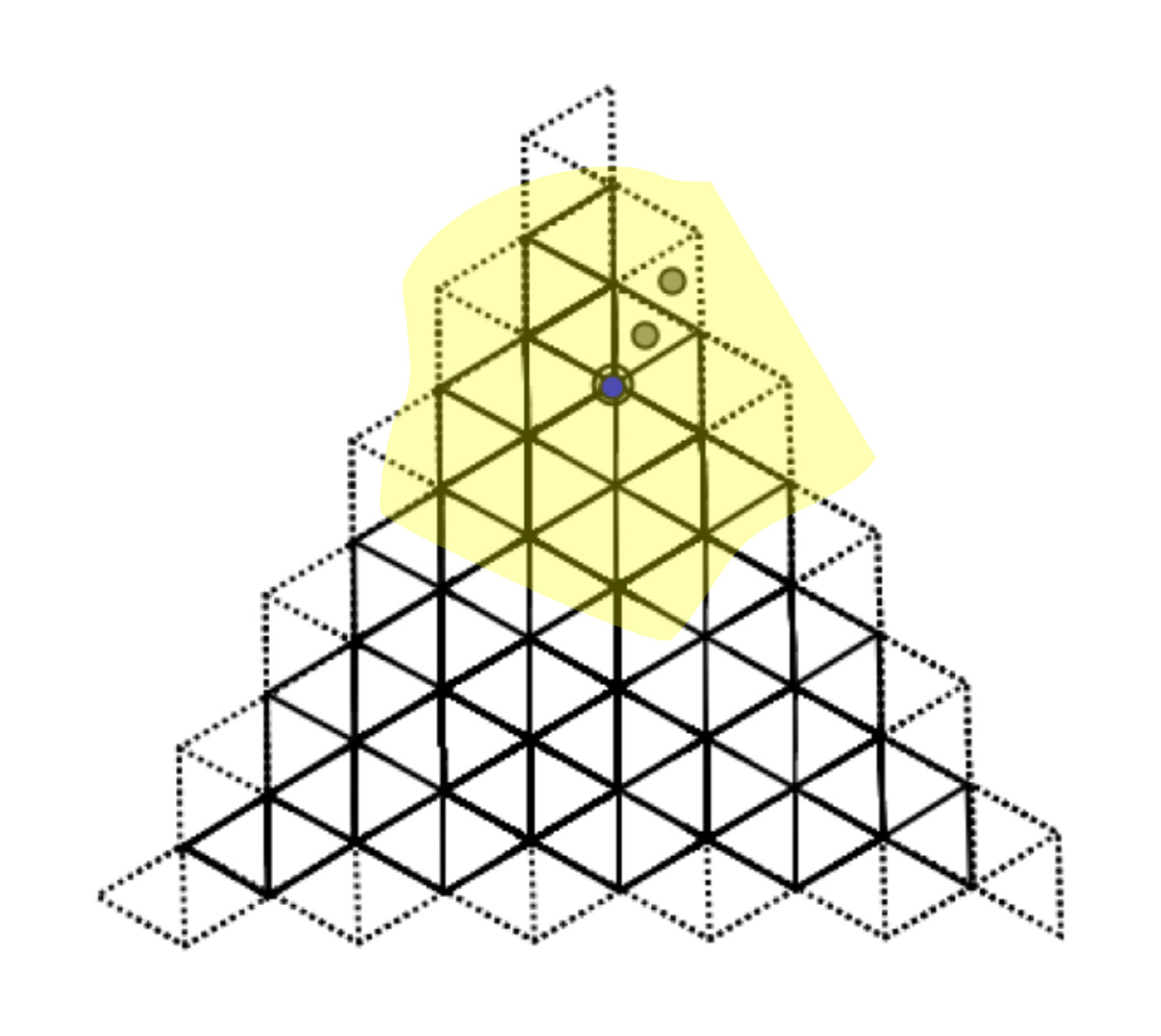} 
\caption{The remaining violated syndromes (in software) after the proposed charge transport corrections from the first step of the GCC~decoder on the errors from Fig.~\ref{fig:GCCStep0}. The remaining combination of initial errors and introduced errors are given in grey.}
\label{fig:GCCStep1}
\end{figure}

Fig.~\ref{fig:GCCStep1} shows the remaining stabilizer violations after the first step of the GCC~algorithm for the discussed example. The introduced correction, due to charge transport, can combine from different colors to cancel, and the remaining error (initial error plus correction) are given in grey. Therefore, since the top-right cluster from Fig.~\ref{fig:GCCStep0} is not charge neutral, there remains a charge after the initial correction, represented by the blue violation in Fig.~\ref{fig:GCCStep1}. At the second step, the charge is close enough (distance~2) to the Blue boundary to cancel out the charge with the boundary, therefore correcting the error locally in the top right section of the graph. As such, the decoder will successfully correct this form of error. However, it should be noted that the growth in the cluster size can very quickly span across the graph, resulting in potential logical error, if there are too many initial non-neutral clusters. This will happen when close to threshold, due to percolation, and as such the threshold of such a scheme is lower than ideal decoders such as that proposed by Delfosse~\cite{PhysRevA.89.012317}.

\end{document}